\documentclass[11pt]{article}
\usepackage{latexsym,amssymb}
\usepackage{epic,eepic}
\newcommand{\SB}{\{\,}
\newcommand{\SM}{\;{:}\;}
\newcommand{\SE}{\,\}}
\newcommand{\Card}[1]{|#1|}
\newcommand{\hy}{\hbox{-}\nobreak\hskip0pt}
\newcommand{\ola}{\mbox{\normalfont ola}}
\newcommand{\prf}{\mbox{\normalfont prf}}
\newcommand{\cost}{\mbox{\normalfont c}}
\newcommand{\ncost}{\mbox{\normalfont nc}}
\newcommand{\NP}{\mbox{\normalfont NP}}
\newenvironment{proof}{\pf}{\qed}

\begin{document}
\newcommand{\2}{\vspace{0.2 cm}}
\newcommand{\dom}{\mbox{$\rightarrow$}}
\newcommand{\ndom}{\mbox{$\not\rightarrow$}}
\newcommand{\sdom}{\mbox{$\Rightarrow$}}
\newcommand{\nsdom}{\mbox{$\not\Rightarrow$}}
\newcommand{\qed}{\hfill$\Box$}
\newcommand{\pf}{{\bf Proof: }}
\newtheorem{theorem}{Theorem}[section]
\newtheorem{algorithm}[theorem]{Algorithm}
\newtheorem{proposition}[theorem]{Proposition}
\newtheorem{lemma}[theorem]{Lemma}
\newtheorem{problem}[theorem]{Problem}
\newtheorem{corollary}[theorem]{Corollary}
\newtheorem{conjecture}[theorem]{Conjecture}
\newtheorem{remark}[theorem]{Remark}
\newcommand{\beq}{\begin{equation}}
\newcommand{\eeq}{\end{equation}}
\newcommand{\ra}{\rangle}
\newcommand{\la}{\langle}
\newcommand{\har}{\rightleftharpoons}
\newcommand{\<}[1]{\mbox{$\la #1 \ra$}}

\newcommand{\RESx}{2.5551}     
 \newcommand{\RESxy}{2.555}    
 \newcommand{\RESz}{5.88}     

 \newcommand{\Ax}{0.119}
 \newcommand{\Bx}{1.96}

 \newcommand{\Xx}{0.967095}
 \newcommand{\Yx}{2}

 \newcommand{\Cx}{1.4625}
 \newcommand{\Qx}{4}

\title{The Linear Arrangement Problem Parameterized  Above Guaranteed Value}

\date{}

\author
{Gregory Gutin\thanks{Corresponding author. Department of Computer
Science, Royal Holloway University of London, Egham, Surrey TW20
OEX, UK, gutin@cs.rhul.ac.uk and Department of Computer Science,
University of Haifa, Israel} \and Arash Rafiey\thanks{Department of
Computer Science, Royal Holloway University of London, Egham, Surrey
TW20 OEX, UK, arash@cs.rhul.ac.uk} \and Stefan Szeider\thanks{
Department of Computer Science, Durham University Science Labs,
South Road, Durham DH1 3LE, UK, stefan.szeider@durham.ac.uk} \and
Anders Yeo\thanks{Department of Computer Science, Royal Holloway
University of London, Egham, Surrey TW20 OEX, UK,
anders@cs.rhul.ac.uk}}

\maketitle

\begin{abstract}
  \noindent
  A linear arrangement (LA) is an assignment of distinct integers to
  the vertices of a graph.  The cost of an LA is the sum of lengths of
  the edges of the graph, where the length of an edge is defined as
  the absolute value of the difference of the integers assigned to its
  ends.  For many application one hopes to find an LA with small cost.
  However, it is a classical NP-complete problem to decide whether a
  given graph $G$ admits an LA of cost bounded by a given integer.  Since
  every edge of $G$ contributes at least one to the cost of any LA, the
  problem becomes trivially fixed-parameter tractable (FPT) if
  parameterized by the upper bound of the cost. Fernau asked whether
  the problem remains FPT if parameterized by the upper bound of the
  cost minus the number of edges of the given graph; thus whether the
  problem is FPT ``parameterized above guaranteed value.''  We answer
  this question positively by deriving an algorithm which decides in
  time $O(m+n+5.88^k)$ whether a given graph with $m$ edges and
  $n$ vertices admits an LA of cost at most $m+k$ (the algorithm
  computes such an LA if it exists).  Our algorithm is based on a
  procedure which generates a problem kernel of linear size in linear
  time for a connected graph $G$. We also prove that more general parameterized LA problems
  stated by Serna and Thilikos are not FPT, unless P=NP.

  {\em Key words:} linear arrangement, fixed-parameter tractability,
  parametrization above guaranteed value, para-NP-complete.
\end{abstract}

\section{Introduction}
All graphs considered in this paper do not have loops or parallel
edges. A \emph{linear arrangement} of a graph $G=(V,E)$ is a
one-to-one mapping $\alpha:V\rightarrow \{1,\dots,\Card{V}\}$. The
\emph{length} of an edge $uv\in E$ relative to $\alpha$ is defined
as
\[
\lambda_\alpha(uv)=|\alpha(u)-\alpha(v)|.
\]
The \emph{cost} $\cost(\alpha,G)$ of a linear arrangement $\alpha$
is the sum of lengths of all edges of $G$ relative to $\alpha$,
i.e.,
\[
\cost(\alpha,G)=\sum_{e\in E} \lambda_\alpha(e).
\]
Linear arrangements of minimal cost are \emph{optimal}; $\ola(G)$
denotes the cost of an optimal linear arrangement of $G$.

The {\bf Linear Arrangement Problem (LAP)} is the problem of
deciding whether, given a graph $G$ and an integer $k$, $G$ admits
a linear arrangement of cost at most $k$, i.e., whether
$\ola(G)\leq k$.  The problem has numerous application; in
particular, the first published work on the subject appears to be
the 1964 paper of Harper~\cite{harperJSIAM12}, where a
polynomial-time algorithm for finding optimal linear arrangement
for $n$-cubes is developed, which has applications in
error-correcting codes.  Goldberg and Klipker \cite{goldberg1976}
were first to obtain a polynomial-time algorithm for computing
optimal linear arrangements of trees. Faster algorithms for trees
were obtained by Shiloach~\cite{shiloachSIAMJC8} and
Chung~\cite{chungCMA10}. However, we cannot hope to find optimal
linear arrangements for the class of all graphs in polynomial time
since  LAP is a classical $\NP$\hy complete problem
\cite{GareyJohnson79,GareyJohnsonStockmeyer76}.

Recently, LAP was studied under the framework of parameterized
complexity \cite{fernau2005,sernaEATCSB86}.  We recall some basic
notions of parameterized complexity here, for a more in-depth
treatment of the topic we refer the reader to
\cite{downey1999,EFLRacid05,fernau2005,flum2006,Niedermeier06}.  A
parameterized problem $\Pi$ can be considered as a set of pairs
$(I,k)$ where $I$ is the \emph{problem instance} and $k$ (usually an
integer) is the \emph{parameter}.  $\Pi$ is called
\emph{fixed-parameter tractable (FPT)} if membership of $(I,k)$ in
$\Pi$ can be decided in time $O(f(k)|I|^c)$, where $|I|$ is the size
of $I$, $f(k)$ is a computable function, and $c$ is a constant
independent from $k$ and $I$. Let $\Pi$ and $\Pi'$ be parameterized
problems with parameters $k$ and $k'$, respectively. An
\emph{fpt-reduction $R$ from $\Pi$ to $\Pi'$} is a many-to-one
transformation from $\Pi$ to $\Pi'$, such that (i) $(I,k)\in \Pi$ if
and only if $(I',k')\in \Pi'$ with $k'\le g(k)$ for a fixed
computable function $g$ and (ii) $R$ is of complexity
$O(f(k)|I|^c)$. A \emph{reduction to problem kernel} (or
\emph{kernelization}) is an fpt-reduction $R$ from a parameterized
problem $\Pi$ to itself. In kernelization, an instance $(I,k)$ is
reduced to another instance $(I',k')$, which is called the
\emph{problem kernel}.
It is easy to see that a decidable parameterized problem is FPT if
and only if it admits a kernelization (see, e.g.,
\cite{EFLRacid05,Niedermeier06}); however, the problem kernels
obtained by this general result have impractically large size.
Therefore, one tries to develop kernelizations that yield problem
kernels of smaller size, if possible of size linear in the
parameter.

The following is a straightforward way to parameterize LAP
\cite{fernau2005,sernaEATCSB86}:

\begin{quote}
  \noindent{\bfseries Parameterized LAP}\\
  \emph{Instance:} A graph $G$.\\
  \emph{Parameter:} A positive integer $k$.\\
  \emph{Question:} Does $G$ have a linear arrangement of cost at most
  $k$?
\end{quote}

An edge has length at least $1$ in any linear arrangement. Thus,
for a graph $G$ with $m$ edges always $\ola(G)\geq m$ prevails; in
other words, $m$ is a guaranteed value for $\ola(G)$.
Consequently, parameterized LAP is FPT by trivial reasons (we
reject a graph with more than $k$ edges and solve LAP by brute
force if the graph has at most $k$ edges).  Hence it makes sense
to consider the \emph{net cost} $\ncost(\alpha,G)$ of a linear
arrangement $\alpha$ defined as follows:
\[
\ncost(\alpha,G)=\sum_{e\in E} (\lambda_\alpha(e)-1)=
\cost(\alpha,G)-m.
\]
We denote the net cost of an optimal linear arrangement of $G$ by
$\ola^+(G)$. Indeed, the following non-trivial parameterization of
LAP is considered by Fernau~\cite{fernau2005,fernauD}:

\begin{quote}
  \noindent{\bfseries LA parameterized above guaranteed value (LAPAGV)}\\
  \emph{Instance:} A graph $G$.\\
  \emph{Parameter:} A positive integer $k$.\\
  \emph{Question:} Does $G$ have a linear arrangement of net cost at
  most $k$?
\end{quote}
Parameterizations above a guaranteed value were first considered by
Mahajan and Raman \cite{MahajanRaman99} for the problems Max-SAT and
Max-Cut; such parameterizations have lately gained much attention
\cite{fernau2005,Niedermeier06}. However, apparently only a few
nontrivial problems parameterized above guaranteed value are known
to be FPT.

Fernau \cite{fernau2005,fernauD,fernauM} raises the question of
whether LAPAGV is FPT (the status of this problem was reported open
in Cesati's compendium \cite{Cesati05}).  We answer this question
positively by deriving a kernelization procedure for LAPAGV that
yields problem kernels of linear size in linear time for connected
graphs $G$. Moreover, using the method of bounded search trees, we
develop an algorithm that solves LAPAGV for the obtained kernel more
efficiently than by brute force.  In summary, we obtain an algorithm
that decides in time $O(m+n+5.88^k)$ whether a given graph with $m$
edges and $n$ vertices admits an LA of cost at most $m+k$. Our
algorithm also produces an optimal linear arrangement if
$\ola^+(G)\le k.$  A key concept of our kernelization is the
suppression of vertices of degree 2, a standard technique used in
the design of parameterized algorithms (e.g., for finding small
feedback vertex sets in graphs \cite{downey1999}). For LAPAGV,
however, we need a more sophisticated approach where we suppress
only vertices of degree 2 that satisfy a certain condition depending
on the parameter~$k$.

Fernau \cite{fernauM} proposes a bounded search tree approach to
prove that LAPAGV is FPT. The description of the approach is
incomplete (for example, it is unclear how to deal with vertices of
degree 2 without rejecting any yes-instances) and an inequality,
which is required by Fernau's approach to show that LAPAGV is FPT,
is not proved. These conclusions are confirmed in our private
communication with Fernau (February, 2006) and it remains to be seen
whether a bounded search tree approach can be used to prove that
LAPAGV is FPT.

Serna and Thilikos \cite{sernaEATCSB86} formulate more general
parameterized LA problems (see Section~\ref{strongsec}) and ask
whether their problems are FPT. We prove that the problems are not
FPT (unless P=NP) by demonstrating that for almost all fixed
values of the parameter, the corresponding decision problems are
NP-complete. This implies that the problems are para-NP-complete
\cite{flum2006}. We conclude the paper by Theorem \ref{st3}, which
indicates that our FPT result cannot be extended much further, in
a sense.

For a graph $G$ and a set $X$ of its vertices, $V(G)$, $E(G)$ and
$G[X]$ denote the vertex set of $G$, the edge set of $G$, and the
subgraph of $G$ induced by $X$, respectively. An edge $e$ in a graph
$G$ is a {\em bridge} if $G-e$ has more components than $G$ has. A
connected graph with at least two vertices and without bridges is
called {\em 2-edge-connected}. A {\em bridgeless component} of a
graph $G$ is a maximal induced subgraph of $G$ with no bridges.
Observe that the bridgeless components of $G$ are the connected
components that we get after removing all bridges from $G.$ A
bridgeless component is either a 2-edge-connected graph or is
isomorphic to $K_1$; in the latter case we call it {\em trivial}.
Further graph-theoretic terminology can be found in Diestel's book
\cite{Diestel00}.

\section{Kernelization}

In the next section, we use the following simple lemma to solve
LAPAGV for the general case of an arbitrary graph input $G$. The
lemma allows us to confine our attention to connected graphs in the
rest of this section.

\begin{lemma}\label{lem:components}
  Let $G_1,\dots,G_p$ be the connected components of a graph $G$.
  Then $\ola^+(G)=\sum_{i=1}^p \ola^+(G_i)$.
\end{lemma}
\begin{proof}
  Follows directly from the definitions.
\end{proof}

\2

Let $\alpha$ be a linear arrangement of a graph $G$. It is
convenient to use for subgraphs $G'$ of $G$ the notation
$\ncost(\alpha,G')=\sum_{uv\in E(G')} (\lambda_{\alpha}(uv) -1)$.

\begin{lemma}\label{lem:jump}
  Let $G$ be a graph, let $X\subseteq V(G)$, and let $u,v$ be two
  distinct vertices of $G$ that belong to the same connected component
  of $G-X$.  Let $\alpha$ be a linear arrangement of $G$ with
  $\alpha(u) < \alpha(x) < \alpha(v)$ for every $x\in X$.  Then
  $\ncost(\alpha,G-X)\geq \Card{X}$.
\end{lemma}
\pf
  We proceed by induction on $\Card{X}$. If $\Card{X}=0$ then the
  lemma holds vacuously. Hence we assume $\Card{X}\geq 1$ and pick
  $x\in X$. We define $G'=G-x$, $X'=X\setminus \{x\}$, and we let
  $\alpha'$ be the linear arrangement of $G'$ obtained from $\alpha$
  by setting, for $y\in V(G')$, $\alpha'(y)=\alpha(y)$ if $\alpha(y)<
  \alpha(x)$, and $\alpha'(y)=\alpha(y)-1$ otherwise.  By induction
  hypothesis, $\ncost(\alpha',G'-X')\geq \Card{X'}$.  By assumption,
  $G-X$ contains a path $P$ from $u$ to $v$; hence $P$ contains at
  least one edge $w_1w_2$ with $\alpha(w_1)<\alpha(x)<\alpha(w_2)$
  (and $w_1,w_2\notin X$).  By definition of $\alpha'$, we have
  $\lambda_\alpha(w_1w_2)=\lambda_{\alpha'}(w_1w_2)+1$.  Since for all other
  edges $e\in E(G'-X')$ we have $\lambda_\alpha(e)\geq\lambda_{\alpha'}(e)$,
  $\ncost(\alpha,G-X)\geq \ncost(\alpha',G'-X') +1$ follows.
\qed

Let $G$ be a connected graph and let $\alpha$ be a linear
arrangement of $G$.  We say that two disjoint subgraphs $A,B$ of
$G$ are \emph{$\alpha$\hy comparable} if either $\alpha(a) <
\alpha(b)$ holds for all $a\in V(A),b\in V(B)$, or $\alpha(a) >
\alpha(b)$ holds for all $a\in V(A),b\in V(B)$.  Moreover, let $e$
be a bridge of $G$ and let $G_1,G_2$ be the two connected
components of $G-e$. For a positive integer $k$, we say that $e$
is \emph{$k$\hy separating} if both $\Card{V(G_1)},\Card{V(G_2)} >
k$.

\begin{lemma}\label{lem:bridges}
Let $G$ be a connected graph and let $k$ be a positive integer such
that $k\geq \ola^+(G)$. Then for every optimal linear arrangement
$\alpha$ of $G$ and every $k$\hy separating bridge $e$ of $G$, the
two connected components of $G-e$ are $\alpha$\hy comparable.
\end{lemma}
\pf Let $\alpha$ be an optimal linear arrangement. Let $e$ be a
$k$\hy separating bridge of $G$ and let $G_1,G_2$ be the two
connected components of $G-e$. Since $e$ is a $k$\hy separating
bridge, $\Card{V(G_1)},\Card{V(G_2)} > k$ holds by definition.  We
denote the extremal values of the vertices of $G_1$ and $G_2$ with
respect to $\alpha$ by $l_i= \min_{v\in V(G_i)}\alpha(v)$ and $r_i=
\max_{v\in V(G_i)} \alpha(v)$, $i=1,2$.  We may assume, w.l.o.g.,
that $l_1<l_2$.

  First we show that $r_1<r_2$.  Assume to the contrary that
  $r_1>r_2$. Now $\alpha^{-1}(l_1)$ and $\alpha^{-1}(r_1)$ belong to
  the same connected component of $G-V(G_2)$, and Lemma~\ref{lem:jump}
  implies $\ncost(\alpha,G)\geq \Card{V(G_2)}> k$, contradicting the
  assumption $\ncost(\alpha,G)\leq k$. Hence indeed $l_1<l_2$ and
  $r_1<r_2$.

  Next we show that $r_1 < l_2$.  Assume to the contrary that $l_2 <
  r_1$.  From $\alpha$ we obtain a new linear arrangement $\alpha'$ of
  $G$, changing the order of vertices in $X=\SB x\in V(G) \SM l_2\leq
  \alpha(x) \leq r_1 \SE$ such that $G_1$ and $G_2$ become
  $\alpha'$\hy comparable, without changing the relative order of
  vertices within $G_1$ or changing the relative order of vertices
  within $G_2$.  That is, for $X\cap
  V(G_i)=\{v^{(i)}_1,\dots,v^{(i)}_{j_i}\}$ and
  $\alpha(v^{(i)}_1)<\dots< \alpha(v^{(i)}_{j_i})$, $i=1,2$, we have
  $\alpha'(v^{(1)}_1)<\dots< \alpha'(v^{(1)}_{j_1}) <
  \alpha'(v^{(2)}_1)<\dots< \alpha'(v^{(2)}_{j_2})$.

  Since $e$ is a bridge, we have
  \begin{equation}\label{e1}
  \ncost(\alpha,G)=\ncost(\alpha,G-e) +
  \lambda_\alpha(e) -1 \mbox{ and }
  \ncost(\alpha',G)=\ncost(\alpha',G-e)+
  \lambda_{\alpha'}(e) -1\end{equation}  Although $\lambda_{\alpha'}(e)$ can be greater
  than $\lambda_{\alpha}(e)$, we will show that an increase of the length
  of $e$ is more than compensated by the reduced cost of $G-e$ under
  $\alpha'$.  Again using Lemma~\ref{lem:jump} we conclude that
  $\ncost(\alpha',G_i)\leq \ncost(\alpha,G_i)-\Card{X\cap V(G_{3-i})}$
  holds for $i=1,2$ (observe that the vertices
  $\alpha^{-1}(l_i),\alpha^{-1}(r_i)$ are in the same component of
  $G-V(G_{3-i})$, and for each vertex $x$ in $X\cap V(G_i)$ we have
  $\alpha(l_i)<\alpha(x) < \alpha(r_i)$).  In summary, we have
  \begin{equation}\label{e2}
    \ncost(\alpha',G-e) \leq \ncost(\alpha,G-e)-\Card{X}.
  \end{equation}

  Using the fact that $|\alpha(x)-\alpha'(x)|\leq \Card{X}-1$ holds
  for all vertices $x\in V(G)$, it is easy to see that
  \begin{equation}\label{e3}
  \lambda_{\alpha'}(e)\leq  \lambda_{\alpha}(e) + \Card{X}-1.
  \end{equation}
  Indeed, if at least one of the ends of $e$ is in $V(G)\setminus X$,
  then clearly $\lambda_{\alpha'}(e)\leq \lambda_{\alpha}(e) + \Card{X}-1$;
  otherwise, if both ends of $e$ are in $X$, then $\lambda_\alpha'(e)\leq
  \Card{X}-1$, and since $\lambda_{\alpha}(e)\geq 1$, we have even
  $\lambda_{\alpha'}(e)\leq \lambda_{\alpha}(e) + \Card{X}-2$.

  By (\ref{e1}),(\ref{e2}) and (\ref{e3}), we obtain $\ncost(\alpha',G)
  \le \ncost(\alpha,G) -1$.  This contradicts the assumption that $\alpha$
  is an optimal linear arrangement.  Hence $l_1 < r_1 < l_2 < r_2$,
  and so $G_1$ and $G_2$ are $\alpha$\hy comparable as claimed.
\qed

\begin{lemma}\label{lem:bridgeless}
  If $G$ is a connected bridgeless graph of order $n\ge 1$, then
  $\ola^+(G)\geq (n-1)/2$.
\end{lemma}
\pf  If $n \leq 2$, then the inequality trivially holds. Thus, we
may assume that $n\ge 3$ and $G$ is 2-edge-connected. Let $\alpha$
be an optimal linear arrangement of $G$ and put $u=\alpha^{-1}(1)$
and $w=\alpha^{-1}(n)$.
 Since $G$ is $2$\hy edge-connected,  Menger's Theorem (see, e.g.,
\cite{Diestel00}) implies that there are two paths $P,P'$ between
$u$ to $w$ such that $E(P)\cap E(P')=\{u,w\}$. Observe that the
subgraph $G'$ of $G$ induced by $E(P)\cup E(P')$ is a collection
of $t\ge 1$ edge-disjoint cycles. Let $n'$ be the number of
vertices in $G'$. Since $G'$ has $t-1$ vertices of degree 4 and
$n'-t+1$ vertices of degree 2, $|E(G')|=(n'-t+1)+2(t-1)=n'+t-1.$
Since $n'\le n$ and $t\le {n-1 \over 2}$, we conclude that
$|E(G')|\le {3 \over 2}(n-1).$ Observe that $\ncost(\alpha,P)\ge
n-1-|E(P)|$ and $\ncost(\alpha,P')\ge n-1-|E(P')|$. Hence,
$$\ola^+(G)=\ncost(\alpha,G)\ge \ncost(\alpha,G')\ge 2(n-1)-|E(G')|\ge
(n-1)/2.$$ \qed

\begin{lemma}\label{vertexcon}
A connected graph $G$ on at least two vertices has a pair $u,v$ of
distinct vertices such that both $G-u$ and $G-v$ are connected.
\end{lemma}
\pf Let $T$ be a spanning tree in $G$ and let $u,v$ be leaves in
$T$. Then $T-x$ is a spanning tree in $G-x$ for $x\in
\{u,v\}.$\qed

\2

 Let $\alpha$ be an optimal linear arrangement of $G$. We call a
vertex $u\in V(G)$ {\em $\alpha$-special} if $G-u$ is connected
and $\alpha(u) \not\in \{1,n\}$.

\begin{lemma} \label{lA2}
Let $G$ be a connected graph. Let $X$ be a vertex set of $G$ such
that $G[X]$ is connected and let $G-X$ have connected components
$G_1,G_2,\ldots ,G_r$ with $n_1,n_2,\ldots ,n_r$ vertices,
respectively, such that $n_1\le n_2\le \ldots \le n_r$. Then
$\ola^+(G) \ge \ola^+(G[X])+\sum_{i=1}^{r-2}n_i.$
\end{lemma}
\pf Let $\alpha$ be an optimal linear arrangement of $G$. If $r\le
2$, then $\sum_{i=1}^{r-2}n_i=0$ and, thus, this lemma holds. Now
assume that $r\ge 3.$ By Lemma \ref{vertexcon}, each nontrivial
$G_i$ has a pair $u_i,v_i$ of distinct vertices such that $G_i-u_i$
and $G_i-v_i$ are connected. If $G_i$ is trivial, i.e., it has just
one vertex $x$, then set $u_i=v_i=x.$ Since $r\ge 3$, for some $j\in
\{1,2,\ldots ,r\}$, we have $\alpha(u_j) \not\in \{1,n\}$ and
$\alpha(v_j) \not\in \{1,n\}$. Now we claim that there is a vertex
$u\in V(G_j)$ such that $G-u$ is connected. Indeed, we set $u=u_j$
if there are edges between $v_j$ and $G[X]$, we set $u=v_j$,
otherwise.

We have proved that $G$ has an $\alpha$-special vertex $u$ not in
$X$. Let $\alpha_u$ be a linear arrangement  of $G-u$ defined as
follows: $\alpha_u(x)=\alpha(x)$ for all $x\in V(G)$ with
$\alpha(x)<\alpha(u)$, and $\alpha_u(x)=\alpha(x)-1$ for all $x\in
V(G)$ with $\alpha(x)>\alpha(u)$. Since $G$ is connected, it has
an edge $yz$ such that $\alpha(y)<\alpha(u)<\alpha(z)$. Observe
that $\lambda_{\alpha}(yz)=\lambda_{\alpha_u}(yz)+1.$ Hence, we
have
$$\ola^+(G)=\ncost(\alpha,G)\ge \ncost(\alpha_u,G-u)+1\ge
\ola^+(G-u)+1.$$ Thus,

\begin{equation}\label{cl}
\ola^+(G)\ge \ola^+(G-u)+1 \mbox{ for an $\alpha$-special vertex }u
\mbox{ of }G
\end{equation}

Run the following procedure: while $G-X$ has a least three
components, choose a $\beta$-special vertex $u\not\in X$ of $G$
for an optimal linear arrangement $\beta$ of $G$ and replace $G$
with $G-u$. By the end of this procedure, we have deleted some $t$
vertices from $G$ obtaining a subgraph $H$ of $G$. By (\ref{cl}),
we have $\ola^+(G)\ge \ola^+(G[X])+t$. Observe that  $H-X$ has at
most two components, if all vertices of at least $r-2$ components
$G_1,G_2,\ldots ,G_r$ are deleted from $G$ during the procedure.
Thus, $t\le \sum_{i=1}^{r-2}n_i$ and $\ola^+(G)\ge \ola^+(G[X])+
\sum_{i=1}^{r-2}n_i.$\qed

\begin{figure}[h!]
\begin{center}
\setlength{\unitlength}{0.021667in}
\begingroup\makeatletter\ifx\SetFigFont\undefined%
\gdef\SetFigFont#1#2#3#4#5{%
  \reset@font\fontsize{#1}{#2pt}%
  \fontfamily{#3}\fontseries{#4}\fontshape{#5}%
  \selectfont}%
\fi\endgroup%
{\renewcommand{\dashlinestretch}{30}
\begin{picture}(203.75,176)(50,20)
\put(63.25,145.25){\oval(37.5,22.5)[]}
\put(119.5,146.75){\oval(39.5,14.5)[]}
\put(181.38,170){\oval(39.25,12)[]}
\put(184.5,145.38){\oval(38.5,13.25)[]}
\put(184.63,122.38){\oval(36.25,10.25)[]}
\put(82,146){\line(1,0){18}}
\multiput(137.5,152)(.047363281,.033691406){512}{\line(1,0){.047363281}}
\put(139.25,147){\line(1,0){26.5}}
\multiput(138.25,142.75)(.054644809,-.033697632){549}{\line(1,0){.054644809}}
\put(61,145.25){\makebox(0,0)[cc]{$F_1$}}
\put(116,146.25){\makebox(0,0)[cc]{$F^*_0$}}
\put(178,169.75){\makebox(0,0)[cc]{$F_2$}}
\put(182.5,145.25){\makebox(0,0)[cc]{$F_3$}}
\put(181.75,122.5){\makebox(0,0)[cc]{$F_4$}}
\put(89.75,149.75){\makebox(0,0)[cc]{$e_1$}}
\put(146.75,164){\makebox(0,0)[cc]{$e_2$}}
\put(152.75,150.5){\makebox(0,0)[cc]{$e_3$}}
\put(152.75,137.25){\makebox(0,0)[cc]{$e_4$}}
\end{picture}
}\vspace{-5cm} \caption{Illustration for Lemma \ref{lA3}.}
\label{ill}
\end{center}
\end{figure}
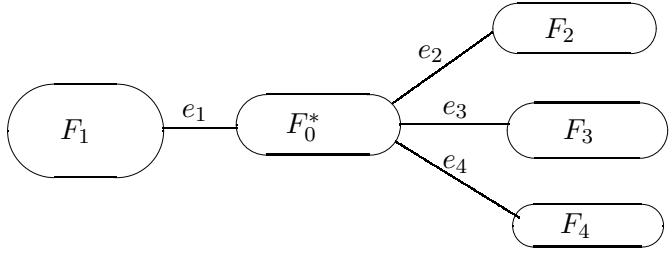

The proof of the next lemma is illustrated in Figure \ref{ill}.

\begin{lemma}\label{lA3}
Let $k$ be a positive integer and let $G$ be a connected graph with
$n$ vertices with $\ola^+(G) \le k$. Then either $G$ has a
$k$-separating bridge or $n \le 4k+1$.
\end{lemma}
\pf Assume that $G$ does not have a $k$-separating bridge. If $G$
is a bridgeless graph, then by Lemma \ref{lem:bridgeless} we know
that $n \le 2k+1$. So, we may assume that $G$ has a bridge. Choose
a bridge $e_1$ with maximal $\min\{\Card{V(F_1)},\Card{V(F_0)}\}$,
where $F_1,F_0$ are the components of $G-e_1$. Assume, w.l.o.g.,
that $\Card{V(F_1)} \leq \Card{V(F_0)}$.  Since $e_1$ is not a
$k$\hy separating bridge, $\Card{V(F_1)} \leq k$ follows of
necessity.  Let $F_0^*$ denote the bridgeless component of $F_0$
that contains a vertex incident to $e_1$. If $F_0=F_0^*$ then
$\Card{V(F_0)}\leq 2k+1$ follows by Lemma~\ref{lem:bridgeless} and
we are done; hence we assume that $F_0\neq F_0^*$.

Let $e_2,\dots,e_r$ denote the bridges of $F_0$ that are incident to
vertices in $F_0^*$. Moreover, let $F_2,\dots,F_r$ denote the
connected components of $F_0-V(F_0^*)$ such that each $e_i$ is
incident with a vertex of $F_i$, $i=2,\dots,r$. Assume that
$|V(F_2)|\ge |V(F_3)|\ge \ldots \ge |V(F_r)|.$ Suppose that
$|V(F_2)|>|V(F_1)|$. Then the component of $G-e_2$ different from
$F_2$ has more vertices than $F_1$, which is impossible by the
choice of $e_1$ and the assumption that $G$ has no $k$\hy separating
bridges. We conclude that $|V(F_1)|\ge |V(F_2)|.$ By Lemma
\ref{lA2}, $\ola^+(G)\ge \ola^+(F_0^*)+\sum_{i=3}^{r}|V(F_i)|.$
Thus, $\sum_{i=3}^{r}|V(F_i)|\le k-\ola^+(F_0^*).$ Since
$|V(F_2)|\le |V(F_1)|\le k$ and, by Lemma~\ref{lem:bridgeless},
$|V(F^*_0)|\le 2\cdot \ola^+(F_0^*)+1$, we obtain that
$$n=|V(F_0^*)|+\sum_{i=1}^{r}|V(F_i)|\le (2\cdot
\ola^+(F_0^*)+1)+(3k-\ola^+(F_0^*))=3k+\ola^+(F_0^*)+1\le
4k+1.$$\qed

\begin{lemma}\label{lA5}
Let $k$ be a positive integer and let $G$ be a connected graph
with the following structure:

\begin{itemize}

\item [1)] $G$ has bridgeless components $C_1,C_2, \ldots ,C_t$,
$t\ge 2$, such that every two consecutive components $C_i$ and
$C_{i+1}$ are linked by a single edge $e_i$, which is a
$k$-separating bridge in $G,$ $i=1,2,\ldots, t-1.$

\item [2)] Let $L=G[\bigcup_{i=1}^tV(C_i)]$. The graph $G'=G-V(L)$
has connected components $G_1,G_2,\ldots ,G_r$ such that each $G_j$
has edges only to one subgraph $C_{\pi(j)}$, $\pi(j)\in \{1,2,\ldots
,t\}.$

\end{itemize}

Let $J_p$ be the indices of all $G_j$ such that $\pi(j)=p,$
$p=1,2,\ldots ,t.$ Let $n_i=\max\{|V(G_j)|:\ j\in J_i\},$
$i=1,2,\ldots ,t.$ Then $\ola^+(G) \ge \ola^+(L)+|V(G')|-n_1-n_t.$
\end{lemma}
\pf Let $\alpha$ be an optimal linear arrangement of $G.$ Let
$$A_p=\left(\bigcup_{j\in J_1\cup J_2 \cup  \cdots \cup J_p} V(G_j)\right)\cup
\left(\bigcup_{j=1}^pV(C_j)\right)$$ for $p=1,2,\ldots ,t.$ By Lemma
\ref{lem:bridges}, the two components of $G-e_1$ are
$\alpha$-comparable. We may assume, w.l.o.g., that
$\alpha(x)<\alpha(y)$ for each $x\in A_1$, $y\not\in A_1$. Because
of the assumption and since the two components of $G-e_2$ are
$\alpha$-comparable, we have $\alpha(x)<\alpha(y)<\alpha(z)$ for
each $x\in A_1$, $y\in A_2-A_1$ and $z\not\in A_2$. Continuing this
argument, we can prove that $\alpha(x_i)<\alpha(x_{i+1})$ for each
$x_i\in A_i$ and $x_{i+1}\in A_{i+1}\setminus \bigcup_{j=1}^{i}A_j.$

By the above conclusion and the arguments similar to those used in
the proof of Lemma \ref{lA2}, we can prove that each $G_j$, apart
from at most one graph $G_p$ with $p\in J_1$ and at most one graph
$G_q$ with $q\in J_t$, has an $\alpha$-special vertex $u$. As in
Lemma \ref{lA2}, it follows that $\ola^+(G-u)\le \ola^+(G)-1.$ Now
we apply a procedure similar to that used in the proof of Lemma
\ref{lA2}:  until $|J_1|\le 1,\ |J_t|\le 1$ and $J_2=\cdots =
J_{t-1}=\emptyset$, choose a $\beta$-special vertex $u\in V(G')$ for
an optimal linear arrangement $\beta$ of $G$ and replace $G$ with
$G-u$ and $G'$ with $G'-u$. The procedure will have at most
$|V(G')|-n_1-n_t$ steps each decreasing $\ola^+(G)$ by at least 1.
Hence $\ola^+(G) \ge \ola^+(L)+|V(G')|-n_1-n_t.$ \qed

\2

Let $G$ be a graph and let $v$ be a vertex of degree 2 of $G$.
Let $vu_1,vu_2$ denote be the edges incident with $v$.  Assume
that $u_1u_2\notin E(G)$.  We obtain a graph $G'$ from $G$ by
removing $v$ (and the edges $vu_1,vu_2$) from $G$ and adding
instead the edge $u_1u_2$. We say that $G'$ is obtained from $G$
by \emph{suppressing} vertex $v$. Furthermore, if the two edges
incident with $v$ are $k$\hy separating bridges for some positive
integer $k$, then we say that $v$ is \emph{$k$\hy suppressible}.
The last definition is justified by the following lemma.

\begin{lemma}\label{lem:suppressing}
Let $G$ be a connected graph and let $v$ be an $\ola^+(G)$\hy
suppressible vertex of $G$.  Then $\ola^+(G)=\ola^+(G')$ holds for
the graph $G'$ obtained from $G$ by suppressing $v$.
\end{lemma}
\pf
  Let $u_1,u_2$ denote the neighbors of $v$ and let $G_1,G_2$ denote
  the connected components of $G-v$ with $u_i\in V(G_i)$, $i=1,2$.
  Consider an optimal linear arrangement $\alpha$ of $G$.  As above we
  use the notation $l_i= \min_{w\in V(G_i)}\alpha(w)$ and $r_i=
  \max_{w\in V(G_i)}\alpha(w)$, $i=1,2$, and we assume, w.l.o.g., that
  $l_1 <l_2$.  Since $vu_1,vu_2$ are $\ola^+(G)$\hy separating
  bridges, Lemma~\ref{lem:bridges} implies that $\alpha$ assigns to
  the vertices of $G_i$ an interval of consecutive integers.  Thus, we conclude that
  $l_1<r_1<\alpha(v)<l_2<r_2$.  We define a linear arrangement
  $\alpha'$ of $G'$ by setting $\alpha'(w)=\alpha(w)$ for $w\in V(G_1)$
  and $\alpha'(w)=\alpha(w)-1$ for $w\in V(G_2)$. Evidently
  $\ola^+(G')\leq \ncost(\alpha',G')=\ncost(\alpha,G)=\ola^+(G)$.

  \sloppypar Conversely, assume that $\alpha'$ is an optimal linear arrangement
  of $G'$.  We proceed symmetrically to the first part of this proof.
  Let $l_i= \min_{w\in V(G_i)}\alpha'(w)$ and $r_i= \max_{w\in
    V(G_i)}\alpha'(w)$, $i=1,2$, and assume, w.l.o.g., that $l_1
  <l_2$.  Observe that $u_1u_2$ is an $\ola^+(G')$\hy separating
  bridge of $G'$, hence Lemma~\ref{lem:bridges} applies. Thus
  $l_1<r_1<l_2<r_2$.  We define a linear arrangement $\alpha$ of $G$
  by setting $\alpha(w)=\alpha'(w)$ for $w\in V(G_1)$,
  $\alpha(v)=r_1+1$, and $\alpha'(w)=\alpha(w)+1$ for $w\in V(G_2)$.
  Evidently $\ola^+(G)\leq
  \ncost(\alpha',G)=\ncost(\alpha',G')=\ola^+(G')$.

  Hence $\ola^+(G)=\ola^+(G')$ as claimed.
\qed

\begin{theorem}\label{the:kernelsize}
Let $k$ be a positive integer, and let $G$ be a connected graph
without $k$\hy suppressible vertices. If $\ola^+(G) \leq k$, then
$G$ has at most $5k+2$ vertices and at most $6k+1$ edges.
\end{theorem}
\pf Let $n=\Card{V(G)}>1$, and let $\ola^+(G)\leq k$.

Any linear arrangement of $G$ can have at most $n-1$ edges of
length 1, and each additional edge contributes at least 1 to the
net cost. Thus, $m\le n-1+k$ and it suffices to show that $n\le
5k+3.$

If $G$ does not have a $k$-separating bridge, then by Lemma
\ref{lA3} we have $n \le 5k+1$. Assume now that $G$ has a
$k$-separating bridge. Let $e=uv$ be such a bridge, and let
$H_1,H_2$ be two connected component of $G-e$, where $H_1$ contains
$u$. Let $C^u$ ($C^v$) be the bridgeless components containing $u$
($v$). Let $C^u_1,C^u_2,..., C^u_p$ ($C^v_1,C^v_2,..., C^v_q$) be
all connected components of $H_1-V(C^u)$ ($H_2-V(C^v)$). Observe
that each of the components $C^u_i$ ($C^v_i$) is linked to $C^u$
($C^v$) by a bridge. Assume that $|V(C^x_i)| \le |V(C^x_j)|$ for
$i<j$, where $x\in \{u,v\}$. By Lemma \ref{lA2}, we have
$\sum_{i=1}^{i=p-1} |V(C^u_i)| \le k$ and $\sum_{i=1}^{i=q-1}
|V(C^v_i)| \le k$. If the bridge between $C^u_p$ and $C^u$ ($C^v_q$
and $C^v$) is $k$-separating, we consider the bridgeless component
of $C^u_p$ ($C^v_q$) containing an endvertex of the bridge and the
connected components obtained from $C^u_p$ ($C^v_q$) by deleting the
vertices of the bridgeless component. Continuation of the procedure
above as long as possible will bring us the following decomposition
of $G$:

\begin{itemize}

\item [1)] $G$ has bridgeless components $C_1,C_2, \ldots ,C_t$,
$t\ge 2$, such that every two consecutive components $C_i$ and
$C_{i+1}$ are linked by a single edge $e_i$, which is a
$k$-separating bridge in $G,$ $i=1,2,\ldots, t-1.$

\item [2)] Let $L=G[\bigcup_{i=1}^tV(C_i)]$. The graph $G'=G-V(L)$
has connected components $G_1,G_2,\ldots ,G_r$ such that each $G_j$
has edges only to one subgraph $C_{\pi(j)}$, $\pi(j)\in \{1,2,\ldots
,t\}.$

\end{itemize}

Since we have carried out the above procedure as long as possible,
all bridges between $G'$ and $L$ are not $k$-separating. Thus,
$|V(G_j)|\le k$ for each $j=1,2,\ldots ,t.$ Recall that $J_p$ is
the set of indices of all $G_j$ such that $\pi(j)=p,$
$p=1,2,\ldots ,t,$ and $n_i=\max\{|V(G_j)|:\ j\in J_i\},$
$p=1,2,\ldots ,t.$ By Lemma \ref{lA5}, $\ola^+(G) \ge
\ola^+(L)+|V(G')|-n_1-n_t.$ Since $n_1\le k,\ n_t\le k$ and
$\ola^+(G)\le k$, we obtain
\begin{equation}\label{eq1}
|V(G')|\le 3k-\ola^+(L).
\end{equation}

Since $G$ has no $k$\hy suppressible vertices, the bridgeless
components $C_2,C_3, \ldots ,C_{t-1}$ are not trivial. Observe
that $\sum_{i=2}^{t-1}\ola^+(C_i)\le \ola^+(L).$ By Lemma
\ref{lem:bridgeless}, every component $\ola^+(C_i)\ge 1$, $2\le
i\le t-1$, and thus $t-2\le \ola^+(L).$ By Lemma
\ref{lem:bridgeless}, $|V(C_i)|\le 2 \cdot \ola^+(C_i)+1$ for each
$i=1,2,\ldots, t.$ Hence,
\begin{equation}\label{eq2}|V(L)|=\sum_{i=1}^t|V(C_i)|\le
2(\sum_{i=1}^t \ola^+(C_i))+t\le 3 \cdot
\ola^+(L)+2.\end{equation} Combining (\ref{eq1}) and (\ref{eq2}),
we obtain
$$|V(G)|=|V(G')|+|V(L)|\le (3k-\ola^+(L))+(3 \cdot \ola^+(L)+2)\le 3k
+ 2\cdot \ola^+(L)+2 \le 5k+2.$$\qed

\begin{theorem}\label{maint}
Let $f(n,m)$ be the time sufficient for checking whether ${\rm
ola}^+(G)\le k$ for a connected graph $G$ with $n$ vertices and $m$
edges. Then $$f(n,m)=O(m+n+f(5k+2,6k+1)).$$
\end{theorem}
\pf We assume that $G$ is represented by adjacency lists. Using a
depth-first-search (DFS) algorithm, we can determine the cut
vertices of $G$ in time $O(n+m)$ (see Tarjan~\cite{Tarjan72}). Let
$T$ be a spanning rooted tree of $G$ (say, as obtained by the DFS
algorithm).  For each vertex $v\in V(G)$, let $T_v$ denote the
subtree of $T$ rooted at $v$. That is, $T_v$ contains $v$ and all
descendants of $v$ in $T$.  We assign to each vertex $v$ the
integer $t_v=\Card{V(T_v)}$. This can be done in time $O(n+m)$ by
a single bottom-up traversal of $T$ where we assign $1$ to leaves,
and to non-leaves we assign the sum of the integers assigned to
their immediate descendants plus one.

Consider now a cut vertex $v$ of $G$ of degree $2$. Let $u,w$ be the
neighbors of $v$. Since the edges $vu$ and $vw$ are bridges of $G$,
they are edges of $T$. It follows now directly from the definition
that $v$ is $k$\hy suppressible if and only if one of the following
conditions holds.

1. $v$ is the root of $T$ and $t_u,t_w>k$.

2. $v$ is not the root of $T$ and $k+1 < t_v < n-k$.

Since these conditions can be checked in constant time for each cut
vertex $v$ of $G$, we can find the set $S$ of all $k$\hy
suppressible vertices of $G$ in time $O(n+m)$.  Note that if $H$ is
the graph obtained by suppressing some $v\in S$, some vertices of
$S\setminus \{v\}$ may not be $k$\hy suppressible in $H$; however,
any $k$\hy suppressible vertex of $H$ belongs to $S\setminus \{v\}$.

We compute a set $S'\subseteq S$ starting with the empty set and
successively adding some of the vertices of $S$ to $S'$.  We visit
the vertices of $G$ according to a bottom-up traversal of $T$ (i.e.,
if $v$ is a descendant of $v'$ then we visit $v$ before $v'$).
During this traversal we assign to each vertex $v$ an integer $t'_v$
which is the number of vertices in $S'\cap V(T_v)$.

Assume we visit a vertex $v\in V(G)\setminus S$. If $v$ is a leaf of
$T$ we put $t'_v=0$; otherwise we let $t'_v$ be the sum of the
values $t'_{v'}$ for the direct descendants $v'$ of $v$ in $T$.
Assume we visit a vertex $v\in S$. Let $u$ and $w$ be the neighbors
of $v$ such that $u$ is a direct descendant of $v$. Let $H$ denote
the graph obtained from $G$ by suppressing all vertices in the
current set $S'$. It follows from the considerations above that $v$
is a $k$\hy suppressible vertex of $H$ if and only if one of the
following conditions holds.

1. $v$ is the root of $T$ and $t_u-t'_u,t_w-t'_w>k$.

2. $v$ is not the root of $T$ and $k+1 < t_v-t'_u <
  n-k-\Card{S'}$.

If $v$ is a $k$\hy suppressible vertex of $H$ we add $v$ to $S'$,
put $t'_v=t'_u+1$, and continue; otherwise we leave $S'$ unchanged,
put $t'_v=t'_u$, and continue.

Performing a further bottom-up traversal of $T$ we suppress the
vertices in $S'$ one after the other, and we are left with a graph
$G'$ which has no $k$\hy suppressible vertices.  If $\Card{V(G')}>
5k+2$ or $|E(G')|>6k+1$, then we know from
Theorem~\ref{the:kernelsize} that $\ola^+(G')>k$. It follows from
Lemma~\ref{lem:suppressing} that $\ola^+(G)>k$ as well, and we can
reject $G$.  On the other hand, if $\Card{V(G')}\leq 5k+2$ and
$|E(G')|\le 6k+1$, then we can find an optimal linear arrangement
$\alpha'$ for $G'$ in time $f(5k+2,6k+1)$.  By means of the
construction in the proof of Lemma~\ref{lem:suppressing} we can
transform in time $O(n+m)$ the arrangement $\alpha'$ into an optimal
linear arrangement $\alpha$ of $G$.\qed

 \2

The proof of Theorem \ref{maint} implies the following:

\begin{corollary}
The problem  LAPAGV, with a connected graph $G$ as an input, has a
linear problem kernel, which can be found in linear time. If
$\ola^+(G)\le k$, then the reduced graph (i.e., kernel) has at
most $5k+2$ vertices and $6k+1$ edges.
\end{corollary}

In the next section, we give an upper bound for the function
$g(k)=f(5k+2,6k+1)$ in Theorem \ref{maint}.

\section{Computing Optimal Linear Arrangements}

\begin{lemma} \label{LemX}
Let $G$ be a $2$-vertex-connected graph on $n$ vertices and let
$\alpha$ be a linear arrangement of $G$. Then $\ncost(\alpha,G) \geq
n-2.$
\end{lemma}
\pf Define $x$ and $y$ such that $\alpha(x)=1$ and $\alpha(y)=n$.
For an edge $e=uv$ in $G$ in which $\alpha(u) < \alpha(v)$, let
$Q(e)=\{ w \mbox{ : } \alpha(u)< \alpha (w) < \alpha(v) \}$. For
every vertex $w \in V(G) - \{x,y\}$, there is a path between $x$ and
$y$ in $G-w$,  and therefore there is an edge $uv$ such that
$\alpha(u)< \alpha (w) < \alpha(v)$. This implies that $\bigcup_{e
\in E(G)} Q(e) = V(G) - \{x,y\}$. Since
$\lambda_{\alpha}(e)-1=|Q(e)|$ we have $$\ncost(\alpha,G) = \sum_{e
\in E(G)} |Q(e)| \geq |\bigcup_{e \in E(G)} Q(e)| = n-2.$$ \qed

\2

Let $n$ and $k$ be nonnegative integers. Let $P_n=p_1 p_2 \ldots
p_n$ be a path of order $n$ and let $OLA_{P_n}^+(k,j)$ be the set of
linear arrangements $\alpha$ of $P_n$ with net cost at most $k$ and
such that $\alpha(p_1)=j$  and $\alpha(p_n)=n$. We will first prove
an upper bound for $|OLA_{P_n}^+(k,j)|$.

\begin{theorem} \label{ThmPN}
For all $n \geq 2$, $ k \geq 0$  and $0 \leq j \leq n-1$, we have
$$ |OLA_{P_n}^+(k,j)| \leq 2^{ \Ax{} n + \Bx{} k - \Xx{} j + \Yx{}} .$$

Furthermore the following holds, when $d_2=0.497534$,

$$ |OLA_{P_n}^+(k,2)| \leq (1-d_2) 2^{ \Ax{} n + \Bx{} k - 2 \cdot \Xx{} + \Yx{}} .$$

\end{theorem}

\pf Let $j>k+1$ and let $G$ be $P_n$ with the extra edge $p_1 p_n$.
By Lemma \ref{LemX}, $\ncost(\alpha,G) \geq n-2$. Since
$\lambda_{\alpha} (p_1 p_n) - 1 = n-j-1$, we conclude that
$\ncost(\alpha,P_n) \geq n-2-(n-j-1) =j-1 > k$. Therefore
$|OLA_{P_n}^+(k,j)|=0$ when $j>k+1$, and the theorem holds in this
case. So assume that $j \leq k+1$.
  We also note that the theorem holds when $k=0$, as in this case
$|OLA_{P_n}^+(k,j)| \leq 2$, so assume that $k \geq 1$.

We will prove the theorem by induction on $n$. Clearly the theorem
is true when $n \leq 4$, as in this case $|OLA_{P_n}^+(k,j)| \leq
(n-2)! \leq 2$. So we may assume that $n>4$.

Let $\alpha \in OLA_{P_n}^+(k,j)$ be arbitrary. Let $\alpha'$ be a
linear arrangement of the path $P_n-p_1$ such that
$\alpha'(z)=\alpha(z)$ when $\alpha(z) < \alpha (p_1)$ and
$\alpha'(z)=\alpha(z)-1$ when $\alpha(z) > \alpha (p_1)$.
Furthermore, let $a=\Ax{}$, $b=\Bx{}$, $x=\Xx{}$, $\Gamma = 2^{ a n
+ bk - x j + \Yx{}}$ and $\gamma=|OLA_{P_n}^+(k,j)|$ and consider
the following two cases:

\2

{\bf Case 1: $j=1$}. Let $\alpha (p_2) =q$. Observe that $\alpha'
\in OLA_{P_{n-1}}^+(k-q+2,q-1)$ since $\alpha' (p_2)= q-1$ and
$\lambda_{\alpha}(p_1 p_2) -1 = q-2$. Since $\alpha'$ is uniquely
determined by $\alpha$ we note that there are at most
$|OLA_{P_{n-1}}^+(k-q+2,q-1)|$ linear arrangements in
$OLA_{P_n}^+(k,j)$ with $\alpha (p_2) =q$. This implies the
following: \hspace{-0.2cm} \begin{eqnarray*}
 \gamma & \leq & \sum_{q=2}^{k+2} |OLA_{P_{n-1}}^+(k-q+2,q-1)| \\
                 & \leq & \left( \sum_{q=2}^{k+2}  2^{ a (n-1) + b (k-q+2) - x (q-1) +\Yx{} } \right) -
                           d_2 2^{ a (n-1) + b (k-3+2) - x (3-1) +\Yx{} } \\
& = & \left(  \left( 2^{-a} \sum_{q=0}^{k} ( 2^{-b-x} )^q \right) - d_2 2^{-a-b-x} \right) \Gamma{} \\
 & \leq &   \left( \frac{2^{-a}}{1-2^{-b-x}} - d_2 2^{-a-b-x} \right) \Gamma{}
\leq   \Gamma{}.
\end{eqnarray*}

{\bf Case 2: $j \geq 2$}. First assume that $q = j - \alpha (p_2)>0
$. Since $P_n-p_1$ is connected, there must be an edge $e$ from the
set of vertices with $\alpha$-values in $\{1,2,\ldots , j-1\}$ to
the set of vertices with $\alpha$-values in $\{j+1,j+2,\ldots ,
n\}$. Observe that $\lambda_{\alpha}(e)=\lambda_{\alpha'}(e)+1$.
Since $\lambda_{\alpha}(p_1p_2)-1=q-1$, the net cost of $\alpha'$ is
at most the net cost of $\alpha$ minus $q$. Since  $\alpha'$ is
uniquely determined by $\alpha$ we note that there are at most
$|OLA_{P_{n-1}}^+(k-q,j-q)|$ linear arrangements in
$OLA_{P_n}^+(k,j)$ with $\alpha (p_2) =j-q$.

Now assume that $q = \alpha (p_2) -j >0 $. Let $p_i$ be the vertex
with $\alpha(p_i)=1$. Observe that the path $p_2 p_3 \ldots p_i$
must contain some edge $e=uv$, where $\alpha(u)>j$ and $\alpha(v)<j$
(as $\alpha(p_2)>j$ and $\alpha(p_i)=1<j$). Furthermore, the path
$p_i p_{i+1} \ldots p_n$ must contain some edge $e'=u'v'$, where
$\alpha(v')>j$ and $\alpha(u')<j$ (as $\alpha(p_n)=n>j$ and
$\alpha(p_i)=1<j$). As above we note that
$\lambda_{\alpha}(e)=\lambda_{\alpha'}(e)+1$ and
$\lambda_{\alpha}(e')=\lambda_{\alpha'}(e')+1$.  Since
$\lambda_{\alpha}(p_1p_2)-1=q-1$, the net cost of $\alpha'$ is at
most the net cost of $\alpha$ minus $q+1$. Since  $\alpha'$ is
uniquely determined by $\alpha$, we note that there are at most
$|OLA_{P_{n-1}}^+(k-q-1,j+q-1)|$ linear arrangements in
$OLA_{P_n}^+(k,j)$ with $\alpha (p_2) =j+q$ (as $\alpha' (p_2)
=j+q-1$). This implies the following when $j \geq 3$ :
\hspace{-1.5cm}
\begin{eqnarray*}
 \gamma & \leq &    \sum_{q=1}^{j-1} |OLA_{P_{n-1}}^+(k-q,j-q)|
                              + \sum_{q=1}^{k-1} |OLA_{P_{n-1}}^+(k-q-1,j+q-1)| \\
& \leq & \sum_{q=1}^{j-1}  2^{ a (n-1) + b(k-q) -x(j-q) +\Yx{} }
                          + \sum_{q=1}^{k-1} 2^{ a (n-1) + b(k-q-1) -x(j+q-1) +\Yx{} }  \\
& = &   \left( 2^{-a-b+x} \sum_{q=0}^{j-2} ( 2^{-b+x} )^q
                          + 2^{-a-2b} \sum_{q=0}^{k-2} (2^{- b-x})^q \right) \Gamma{} \\
& \leq &   \left( \frac{2^{-a-b+x}}{1-2^{-b+x}}  +
                                   \frac{2^{-a-2b}}{1-2^{-b-x}}  \right) \Gamma{}  \leq
                                   \Gamma{}.
\end{eqnarray*}

  When $j=2$ we get the following analogously to above.

\begin{eqnarray*}
 \gamma & \leq &    \sum_{q=1}^{j-1} |OLA_{P_{n-1}}^+(k-q,j-q)|
                              + \sum_{q=1}^{k-1} |OLA_{P_{n-1}}^+(k-q-1,j+q-1)| \\
& \leq &   \left( 2^{-a-b+x} \sum_{q=0}^{2-2} ( 2^{-b+x} )^q
                          + 2^{-a-2b} \sum_{q=0}^{k-2} (2^{- b-x})^q \right) \Gamma{}
              -d_2 2^{a(n-1)+b(k-1-1)-2x} \\
& \leq &    \left( 2^{-a-b+x} + \frac{2^{-a-2b}}{1-2^{-b-x}} -d_2 2^{-a-2b} \right) \Gamma{} \\
& \leq & (1-d_2) \Gamma{} \\
\end{eqnarray*}

This completes the induction proof. \qed

\begin{remark}
Note that Theorem \ref{ThmPN} implies that $|OLA_{P_{5k}}^+(k,1)|=
O(2^{\RESxy{}k}) = O(\RESz{}^k)$. It is possible to prove that
$|OLA_{P_{5k}}^+(k,1)| = \Omega(5.36^k)$, which shows that our
result cannot be significantly improved, in a sense.  Due to space
considerations we do not include the proof of $|OLA_{P_{5k}}^+(k,1)|
= \Omega(5.36^k)$.
\end{remark}

Let $n$ and $k$ be nonnegative integers. Let ${\cal T}_n$ be the set
of trees with $n$ vertices. Let $T\in  {\cal T}_n$ and let $X
\subseteq V(T)$ be arbitrary. Let $OLA_T^+(n,k,X)$ be the set of
linear arrangements $\alpha$ of $T$ with net cost at most $k$ and
such that $\alpha(x) \in \{1,n\}$ for all $x \in X$. Note that
$OLA^+_T(n,k,X)=\emptyset$ if $|X| \geq 3$. Now define $t(n,k,i)$ as
follows:
$$ t(n,k,i) = \max \{ |OLA^+_T(n,k,X)| \mbox{ : } \ T\in  {\cal T}_n, |X|=i \}. $$
In other words, no tree $T$ of order $n$ has more than $t(n,k,i)$
linear arrangements such that the net cost is at most $k$ and $i$
prescribed vertices have to be mapped to either $1$ or $n$ (and
$t(n,k,i)$ is the minimum such value).

For a connected graph $G$, let $T_G$ be a spanning tree of $G$.
Since $\ola^+(T_G) \leq \ola^+(G)$ we only have to check all linear
arrangements in $OLA^+_{T_G}(n,k,\emptyset)$ (but still considering
all edges in $G$ and not just $T_G$) to decide whether $\ola^+(G)
\leq k$. Since $|OLA^+_{T_G}(n,k,\emptyset)| \leq t(n,k,0)$ the
values of $t(n,k,i)$ are of interest (especially when $i=0$). We
will prove an upper bound for $t(n,k,i)$ before indicating how to
generate all linear arrangements in $OLA^+_{T_G}(n,k,\emptyset)$.
Note that $t(n,k,3)=0.$

\begin{theorem} \label{thmX}
For all $n \geq 2$, $ k \geq 0$  and $0 \leq i \leq 3$, we have the
following upper bound:
$$ t(n,k,i) \leq 2^{ \Ax{} n + \Bx{} k - \Cx i+ \Qx{} } . $$
\end{theorem}
\pf We will prove the theorem by induction on $n+k-i$. Clearly the
theorem is true when $n=2$ and $0 \leq i \leq 3$, as in this case
$t(n,k,i)=2$ if $i \in \{0,1,2\}$ and $t(n,k,3)=0$. Furthermore when
$i=3$ the theorem also holds. So now let $i \leq 2$ and $n \geq 3$
(and $k \geq 0$) and assume that the theorem holds for all smaller
values of $n+k-i$.

Let $T$ be a tree of order $n$ and let $X$ be a set of $i$ vertices
in $T$. Let $x$ be a leaf in the tree $T$ and let $y$ be the unique
neighbor of $x$ in $T$. Furthermore if some leaf in the tree $T$
does not belong to $X$ then let $x$ be such a vertex (that is $x
\not\in X$).
 Let $\alpha$ be a linear arrangement of $T$
with net cost at most $k$ and with all vertices $q \in X$ having
$\alpha (q) \in \{1,n\}$.   Let $\alpha'$ be a linear arrangement of
the tree $T-x$ such that $\alpha'(z)=\alpha(z)$ when $\alpha(z) <
\alpha (x)$ and $\alpha'(z)=\alpha(z)-1$ when $\alpha(z) > \alpha
(x)$. Furthermore let $a=\Ax{}$, $b=\Bx{}$, $c=\Cx{}$, $\Gamma = 2^{
a n + bk - c i +4}$, and $\gamma=|OLA_T(n,k,X)|$. Consider the
following three cases:

\2

{\bf Case 1: $x,y \not\in X$.} Observe that there are at most
$t(n,k,i+1)$ linear arrangements $\alpha$ in which $\alpha (x) \in
\{1,n\}$, as we may add $x$ to $X$ and use our induction hypothesis.
So now assume that $\alpha (x) \not\in \{1,n\}.$ Assume that
$\alpha(x) - \alpha(y)=j$. This means that
$\lambda_{\alpha}(xy)-1=j-1$. However, since $\alpha (x) \not\in
\{1,n\}$ and $T-x$ is connected, there must be an edge $e$ from the
set of vertices with $\alpha$-values in $\{1,2,\ldots ,
\alpha(x)-1\}$ to the set of vertices with $\alpha$-values in
$\{\alpha(x)+1,\alpha(x)+2,\ldots , n\}$. Observe that
$\lambda_{\alpha}(e)=\lambda_{\alpha'}(e)+1.$ Therefore, the net
cost of $\alpha'$ is at most the net cost of $\alpha$ minus $j$.
Thus, there are at most $t(n-1,k-j,i)$ linear arrangements
$\alpha'$. Since $\alpha'$ is uniquely determined by $\alpha$ we
note that there are at most $t(n-1,k-j,i)$ linear arrangements
$\alpha$ with $\alpha(x) - \alpha(y)=j$. Analogously, there are at
most $t(n-1,k-j,i)$ linear arrangements $\alpha$ with $\alpha(x) -
\alpha(y)=-j$. The above arguments imply the following:
 \hspace{-0.2cm} \begin{eqnarray*}
  \gamma& \leq & t(n,k,i+1)+ 2 \sum_{j=1}^k t(n-1,k-j,i) \\
                 & \leq & 2^{ a n + b k - c (i+1)+\Qx{}} +
                           2 \sum_{j=1}^k  2^{ a (n-1) + b (k-j) - c i +\Qx{}} \\
& = &   \left( \frac{1}{2^{c}}
                        + \frac{2}{2^{a+b}} \sum_{j=0}^{k-1} ( 2^{- b} )^j
                        \right)\Gamma{} \\
& \leq &   \left( \frac{1}{2^{c}}
                        + \frac{2}{2^{a+b} (1-2^{- b}) }
                        \right)\Gamma{} \leq   \Gamma{}.
\end{eqnarray*}

{\bf Case 2: $x \not\in X$ and $y \in X$.} As in our first case
there are at most $t(n,k,i+1)$ linear arrangements $\alpha$ with
$\alpha(x) \in \{1,n\} $. Now assume that $\alpha(x) \not\in
\{1,n\}$ and assume that $|\alpha(x) - \alpha(y)|=j$, which implies
that $\lambda_{\alpha}(xy)-1=j-1$. As in our first case we observe
that there is an edge $e$ in $T-x$ such that
$\lambda_{\alpha}(e)=\lambda_{\alpha'}(e)+1.$ Therefore, there are
at most $t(n-1,k-j,i)$ linear arrangements $\alpha$ with the above
property. Thus, $\gamma \leq t(n,k,i+1)+ \sum_{j=1}^k t(n-1,k-j,i)$.
By the computations in our first case, this implies that $\gamma
\leq \Gamma$, so we have now proved the case when $x \not\in X$ and
$y \in X$.

\2

{\bf Case 3: $x \in X$.} Since  $|X| \leq 2$ (and $n\geq 3$) we note
that the tree $T$ only has two leaves, by our definition of $x$.
Furthermore $X$ contains both leaves in $T$, which implies that
$T=P_n=p_1 p_2 \ldots p_n$ is a path of order $n$ and
$X=\{p_1,p_n\}$. By Theorem \ref{ThmPN} we now obtain the following:
  $$\gamma \leq  |OLA_{P_n}^+(k,1)| \leq 2^{ \Ax{} n + \Bx{} k - \Xx{} + \Yx{}}
\leq 2^{ \Ax{} n + \Bx{} k - 2 \cdot \Cx{} + \Qx{}} =  \Gamma .$$
We have now bounded the value of $t(n,k,i)$ for all the values we
needed. \qed

\2

\begin{remark}
The values $a=\Ax{}$ and $b=\Bx{}$ in the above proofs could be
changed in such a way that we decrease $a$ but increase $b$ (and
change $c$ and $x$ accordingly) or we could decrease $b$ but
increase $a$ (and change $c$ and $x$ accordingly). However the
values we have chosen are the ones that minimize $5a+b$, as our
final bound is basically $O((2^{5a+b})^k)$.
\end{remark}

It is not difficult to turn the computations in the proof of Theorem
\ref{ThmPN} and Theorem \ref{thmX} into a recursive algorithm that
generates $OLA^+_{T'}(n',k',X')$ for all the relevant $n',k',X'$ and
subtrees $T'$ of $T_G$ and $OLA_{P_{n'}}^+(k',j')$ for all relevant
$n'$, $k'$ and $n'$. After computing $OLA^+_{T_G}(n,k,\emptyset)$ we
only need to calculate the net cost of each linear arrangement in
$OLA^+_{T_G}(n,k,\emptyset)$ with respect to $G$. This way we can
find the value $\ola^+(G)$ if $\ola^+(G)\le k$.

In order to do the above we need to generate at most $(n+1)^2 (n-1)
(k+1)$ sets $OLA^+_{T'}(n',k',X'),$ (there are at most $(n+1)^2$
sets $|X'|$, $2\le n'\le n$ and $0\le k'\le k$). We also need to
generate at most  $n(k+1)(k+2)$ sets $OLA_{P_{n'}}^+(k',j')$ (as $1
\leq n' \leq n$, $0 \leq k' \leq k$ and $0 \leq j \leq k+1$). Each
of the above sets can be computed in at most $n\cdot
t(n,k,\emptyset)$ time (as every set will be of size at most
$t(n,k,\emptyset)$). Thus, we can obtain
$OLA^+_{T_G}(n,k,\emptyset)$ in $O(n(n^3 k + nk^2)
t(n,k,\emptyset))$ time. We then need $O((n+m) t(n,k,\emptyset))$
time to consider each linear arrangement $\alpha$ in
$OLA^+_{T_G}(n,k,\emptyset)$ and compute $\ncost(\alpha,G)$, where
$m=|E(G)|$. So the total time complexity, when $n \leq 5k+2$, is at
most
$$
 O(k^5 t(n,k,\emptyset)) = O(k^5  2^{\Ax{} (5k+2)
 + \Bx{} k }) = O(2^{(5 \cdot \Ax{}+\Bx{}+0.0001)k})
 = O(2^{\RESx{}k}).
$$
We have proved the following:

\begin{theorem}\label{kerneltime}
Let $n$ be the number of vertices in a connected graph $G$ and let
$k$ be a nonnegative integer. If $n\le 5k+2$, then we can check
whether $\ola^+(G)\le k$ and compute $\ola^+(G)$, provided
$\ola^+(G)\le k$, in time $O(2^{\RESx{}k}).$
\end{theorem}

Now we are ready to prove the main result of this paper.

\begin{theorem}\label{main}
Let $G=(V,E)$ be a graph and let $k$ be a nonnegative integer. We
can check whether $\ola^+(G)\le k$ and compute $\ola^+(G)$ provided
$\ola^+(G)\le k$ in time $O(|V|+|E|+\RESz{}^k).$
\end{theorem}
\pf Let $G_1,G_2,\ldots, G_p$ be the connected components of $G$. We
can check, in time $O(|V(G_i)|)$, whether $\ola^+(G_i)=0$ since
$\ola^+(G_i)=0$ if and only if $G_i$ is a path. Thus, in time
$O(|V|)$, we can detect all components of $G$ of net cost zero. By
Lemma \ref{lem:components}, we do not need to take these components
into consideration when computing $\ola^+(G)$. Thus, we may assume
that for all components $G_i$, $i=1,2,\ldots , p$, we have
$\ola^+(G_i)\ge 1$. Thus, if $\ola^+(G)\leq k$, then $\ola^+(G_i)\le
k-p+1.$  By Lemma \ref{lem:components}, Theorems \ref{maint} and
\ref{kerneltime}, and the fact that $\ola^+(G_i)\le k-p+1$ if
$\ola^+(G)\leq k$, we can check whether $\ola^+(G)\le k$ and compute
$\ola^+(G)$ provided $\ola^+(G)\le k$ in time
$O(\sum_{i=1}^p(|V(G_i)|+|E(G_i)|)+p2^{\RESx{}(k-p+1)})=O(|V|+|E|+\RESz{}^k).$
\qed

\section{Stronger Parameterizations of LAP}\label{strongsec}

Serna and Thilikos \cite{sernaEATCSB86} introduce the following
related problems. They ask whether either problem is FPT.

\begin{quote}
  \noindent{\bfseries Vertex Average Min Linear Arrangement (VAMLA)}\\
  \emph{Instance:} A graph $G$.\\
  \emph{Parameter:} A positive integer $k$.\\
  \emph{Question:} Does $G$ have a linear arrangement of cost at
  most $k|V(G)|$?
\end{quote}

\begin{quote}
  \noindent{\bfseries Edge Average Min Linear Arrangement (EAMLA)}\\
  \emph{Instance:} A graph $G$.\\
  \emph{Parameter:} A positive integer $k$.\\
  \emph{Question:} Does $G$ have a linear arrangement of cost at
  most $k|E(G)|$?
\end{quote}

Both problems are not FPT (unless P=NP), which follows from the next
two theorems.

\begin{theorem}\label{st1} For any fixed integer $k\ge 2$, it is $\NP$-complete to decide
whether $\ola(H) \leq k|V(H)|$ for a given graph $H$.
\end{theorem}
\begin{proof}   Let $G$ be a graph and let $r$ be an integer. We know that it
is NP-complete to decide whether $\ola(G) \leq r$ (LAP). Let
$n=|V(G)|$. Let $k$ be a fixed integer, $k\ge 2$. Define $G'$ as
follows: $G'$ contains $k$ copies of $G$, $j$ isolated vertices and
a clique with $i$ vertices (all of these subgraphs of $G'$ are
vertex disjoint). We have $n'=|V(G')|=kn+i + j.$

By the definition of $G'$ and the fact that $\ola(K_i)={i+1 \choose
3}$, we have
\[k\cdot \ola(G)=\ola(G')-\ola(K_i)
                         =\ola(G')-{i+1 \choose 3}.\]
Therefore, $\ola(G) \leq r$ if and only if $\ola(G') \leq kr+{i+1
\choose 3}.$ If there is a positive integer $i$ such that $kr+{i+1
\choose 3}=kn'$ and the number of vertices in $G'$ is bounded from
above by a polynomial in $n$, then $G'$ provides a reduction from
LAP to VAMLA with the fixed $k.$ Observe that $kr+{i+1 \choose 3}\ge
k(kn+i)$ for  $i=6kn$. Thus, by setting $i=6kn$ and $j=r+{1 \over k}
{i+1 \choose 3}-kn- i$, we ensure that $G'$ exists and the number of
vertices in $G'$ is bounded from above by a polynomial in $n$.
\end{proof}

\2

The proof of the following theorem is similar, but  $G'$ is defined
differently: $G'$ contains $k$ copies of $G$, a path with $j$ edges
and a clique with $i$ vertices (all of these subgraphs of $G'$ are
vertex disjoint).

\begin{theorem}\label{st2} For any fixed integer $k\ge 2$, it is $\NP$-complete to decide
whether $\ola(H) \leq k|E(H)|$ for a given graph $H$.
\end{theorem}

For a vertex $v$ in a graph $G=(V,E)$, its {\em closed neighborhood}
$N[v]=\{u\in V:\ uv\in E\}\cup \{v\}.$ The {\em profile} of a linear
arrangement $\alpha$ of $G$  is
$$\prf(\alpha,G) = \sum_{z \in V} (\alpha(z) - \min\{ \alpha(w)
\mbox{ : } w \in N[z] \}).$$ Serna and Thilikos \cite{sernaEATCSB86}
introduce also the following problem and ask whether it is FPT.

\begin{quote}
  \noindent{\bfseries Vertex Average Profile} (VAP)\\
  \emph{Instance:} A graph $G=(V,E)$.\\
  \emph{Parameter:} A positive integer $k$.\\
  \emph{Question:} Does $G$ have a linear arrangement of profile  $\le k|V|$?
\end{quote}

Similarly to Theorem \ref{st1} we can prove that VAP is NP-complete
for every fixed $k\ge 2.$

Recently, Flum and Grohe \cite{flumIC187,flum2006} introduced
para-NP and other parameterized complexity classes. Recall that a
parameterized problem $\Pi$ can be considered as a set of pairs
$(I,k)$ where $I$ is the problem instance and $k$  is the
parameter. $\Pi$ is in \emph{para-NP} if membership of $(I,k)$ in
$\Pi$ can be decided in nondeterministic time $O(f(k)|I|^c)$,
where $|I|$ is the size of $I$, $f(k)$ is a computable function,
and $c$ is a constant independent from $k$ and $I$. Here,
nondeterministic time means that we can use nondeterministic
Turing machine. A parameterized problem $\Pi'$ is {\em
para-NP-complete} if it is in para-NP and for any parameterized
problem $\Pi$ in para-NP there is an fpt-reduction from $\Pi$ to
$\Pi'$. Observe that VAMLA, EAMLA and VAP are in para-NP.
Moreover, it follows directly form our results that the three
problems are para-NP-complete (see Corollary 2.16 in
\cite{flum2006}).

Similarly to Theorem \ref{st2} we can prove the following:

\begin{theorem}\label{st3} For each fixed
$0<\epsilon\le 1$, it is $\NP$-complete to decide whether
$\ola^+(H)\le |E(H)|^{\epsilon}$  for a given graph $H$.
\end{theorem}

Notice that Theorem \ref{main} implies that we can decide, in
polynomial time, whether $\ola(H) \leq |E(H)|+\log |E(H)|$ for a
graph $H$. Theorem \ref{st3} indicates that the possibility to
strengthen the last result is rather limited. It would be
interesting to determine the complexity of the problem to verify
whether $\ola(H) \leq |E(H)|+\log^2 |E(H)|$ for a graph~$H$.

\2

{\bf Acknowledgements} Research of Gutin and Rafiey was supported
in part by the IST Programme of the European Community, under the
PASCAL Network of Excellence, IST-2002-506778. Part of the paper
was written when Szeider was vising Department of Computer
Science, Royal Holloway, University of London.

\end{document}